\documentclass[12pt]{article}
\usepackage{amssymb,amsfonts}
\usepackage{epsf,epsfig}
\begin{document}
\baselineskip 18pt

\title{Low- and high-field induced uniform and staggered magnetizations of a spin ladder with DM term}
\author{P.~N.~Bibikov\\ \it Sankt-Petersburg State University}

\maketitle

\vskip5mm

\begin{abstract}
Analytic expressions for uniform and staggered magnetizations of a
spin ladder with a staggered Dzyaloshinskii-Moriya interaction
along rungs are obtained in the lowest perturbative orders. The
obtained formulas describe magnetic behavior in two marginal
regions related to low ($h\ll h_c$) and high ($h\gg h_s$) magnetic
fields.
\end{abstract}

\section{Introduction}

Recently a spin ladder model with staggered Dzyaloshinskii-Moriya
(DM) term was suggested for explanation of two magnetic anomalies
in dimer system ${\rm Cu}_2{\rm(C}_5{\rm H}_{12}{\rm
N}_{2}{\rm)}_2{\rm Cl}_4$ \cite{1},\cite{2},\cite{3}. The first
anomaly is an appearance of a staggered magnetization in a uniform
magnetic field detected by NMR \cite{1}. The second one is a
pronounced smooth behavior of zero temperature magnetization curve
near the critical $h_c$ and saturation $h_s$ points \cite{2}. A
numerical calculation of these effects within the suggested model
produced a good agreement between the theory and experiment
\cite{3}.

In the present paper we develop an analytical approach to both
these problems studying the DM term perturbatively and considering
a spin ladder in two different regimes related to vicinities of
rung-dimerized and full-polarized ground states. The former
corresponds to low magnetic fields $h\ll h_c$. In this case we put
the special condition on the coupling parameters \cite{4},\cite{5}
in order to begin with exact rung-dimerized ground state of the
nonperturbed Hamiltonian. In the latter regime related to high
fields $h\gg h_s$ the unperturbed system is always fully
polarized. Therefore in this case any additional restriction on
the coupling parameters is unnecessary.

\section{Hamiltonian and ground states of a spin ladder}

We shall study the following Hamiltonian,
\begin{equation}
\hat H=\sum_{n=-\infty}^{\infty}H^0_{n,n+1}+H^{DM}_n,
\end{equation}
where
\begin{equation}
H^0_{n,n+1}=H^{rung}_{n,n+1}+H^{leg}_{n,n+1}+H^{frust}_{n,n+1}+H^{cyc}_{n,n+1}+H^{Zeeman}_{n,n+1}+J_{norm}I,
\end{equation}
and
\begin{eqnarray}
H^{rung}_{n,n+1}&=&
\frac{J_{\bot}}{2}({\bf S}_{1,n}\cdot{\bf S}_{2,n}+{\bf S}_{1,n+1}\cdot{\bf S}_{2,n+1}),\nonumber\\
H^{leg}_{n,n+1}&=& J_{\|}({\bf S}_{1,n}\cdot{\bf S}_{1,n+1}+{\bf
S}_{2,n}\cdot{\bf
S}_{2,n+1}),\nonumber\\
H^{frust}_{n,n+1}&=&J_{frust}({\bf S}_{1,n}\cdot{\bf
S}_{2,n+1}+{\bf S}_{2,n}\cdot{\bf
S}_{1,n+1}),\nonumber\\
H^{cyc}_{n,n+1}&=&J_c(({\bf S}_{1,n}\cdot{\bf S}_{1,n+1})({\bf
S}_{2,n}\cdot{\bf S}_{2,n+1})+({\bf S}_{1,n}\cdot{\bf
S}_{2,n})({\bf S}_{1,n+1}\cdot{\bf S}_{2,n+1})\nonumber\\
&-&({\bf S}_{1,n}\cdot{\bf S}_{2,n+1})({\bf S}_{2,n}\cdot{\bf
S}_{1,n+1})),\nonumber\\
H^{Zeeman}_{n,n+1}&=&-\frac{g\mu_Bh}{2}({\bf S}^z_{1,n}+{\bf
S}^z_{2,n}+{\bf S}^z_{1,n+1}+{\bf
S}^z_{2,n+1}),\nonumber\\
H^{DM}_n&=&(-1)^n{\bf D}\cdot[{\bf S}_{1,n}\times{\bf S}_{2,n}].
\end{eqnarray}
Here ${\bf S}_{i,n}$ ($i=1,2$) are spin-1/2 operators associated
with $n$-th rung while $I$ is an identity matrix. The auxiliary
term $J_{norm}I$ in (2) is need only for zero normalization of the
ground state energy. The vector ${\bf
D}=D(0,\cos\theta,\sin\theta)$ lies in the $y-z$ plane.

When the coupling parameters of $H^0_{n,n+1}$ satisfy a system of
rung-dimerization conditions  \cite{4},\cite{5},
\begin{eqnarray}
J_{frust}&=&J_{\|}-\frac{1}{2}J_c,\quad
J_{norm}=\frac{3}{4}J_{\bot}-\frac{9}{16}J_c,\nonumber\\
J_{\bot}&>&2J_{\|},\quad J_{\bot}>\frac{5}{2}J_c,\quad
J_{\bot}+J_{||}>\frac{3}{4}J_c.
\end{eqnarray}
the ground state related to $H^0_{n,n+1}$ has zero energy and the
rung-dimerized form,
\begin{equation}
|0\rangle_{r-d}=\prod_n|0\rangle_n.
\end{equation}
Here each $|0\rangle_n$ is the singlet state (rung-dimer) related
to $n$-th rung. The related one-particle excitation (often called
a {\it triplon}) is a coherent superposition of excited rungs,
\begin{equation}
|k,tripl\rangle^j=\frac{1}{\sqrt{N}}\sum_n{\rm
e}^{ikn}...|1\rangle^j_n...,\quad j=-1,0,1,
\end{equation}
where $N$ is the number of rungs and "..." denotes a product of
rung-dimers. The corresponding dispersion is the following
\cite{4},
\begin{equation}
E_{tripl}(k,j)=J_{\bot}-\frac{3}{2}J_c-jg\mu_Bh+J_c\cos{k}.
\end{equation}

When the external magnetic field satisfy the system of saturation
conditions,
\begin{eqnarray}
g\mu_Bh&>&2(J_{\|}+J_{frust}),\quad
g\mu_Bh>J_{\bot}+2J_{frust},\quad
g\mu_Bh>J_{\bot}+2J_{\|}+J_c,\nonumber\\
g\mu_Bh&>&\frac{J_{\bot}}{2}+J_{\|}+J_{frust}-\frac{J_c}{4}\nonumber\\
&+&\frac{1}{2}\sqrt{\Big(J_{\bot}-J_{\|}-J_{frust}-\frac{J_c}{2}\Big)^2+
3\Big(J_{\|}-J_{frust}-\frac{J_c}{2}\Big)^2},
\end{eqnarray}
and $J_{norm}=g\mu_Bh-J_{\bot}/4-J_{\|}/2-J_{frust}/2-J_c/16$, the
fully polarized state,
\begin{equation}
|0\rangle_{sat}=\prod_n|1\rangle^1_n,
\end{equation}
is a zero energy ground state for $\hat H^0$.

From the local formula,
\begin{equation}
H^{DM}_n|1\rangle^+_n=(-1)^n\frac{D}{\sqrt{8}}\cos\theta|0\rangle_n,
\end{equation}
one can conclude that the lowest order with respect to $D$
correction to $|0\rangle_{sat}$ originates from the following
branch of coherent excitations (by analogy with (6) we call them
{\it singlons}),
\begin{equation}
|k,singl\rangle=\frac{1}{\sqrt{N}}\sum_n{\rm
e}^{ikn}...|0\rangle_n...
\end{equation}
Unlike (6) here "..." denotes a product of polarized rung triplets
(with $j=1$).

The corresponding dispersion,
\begin{equation}
E_{singl}(k)=g\mu_Bh-J_{\bot}-J_{\|}-J_{frust}-\frac{J_c}{2}+\Big(\frac{J_c}{2}+J_{\|}-J_{frust}\Big)\cos{k},
\end{equation}
may be easily obtained from the $\rm Shr\ddot odinger$ equation
and the following local formulas,
\begin{eqnarray}
H^0_{n,n+1}|0\rangle_n|1\rangle^+_{n+1}&=&\Big(\frac{g\mu_Bh}{2}-\frac{J_{\bot}}{2}-\frac{J_{\|}}{2}-\frac{J_{frust}}{2}-
\frac{J_c}{4}\Big)|0\rangle_n|1\rangle^+_{n+1}\nonumber\\
&+&\Big(\frac{J_c}{4}+\frac{J_{\|}}{2}-\frac{J_{frust}}{2}\Big)|1\rangle^+_n|0\rangle_{n+1},\nonumber\\
H^0_{n,n+1}|1\rangle^+_n|0\rangle_{n+1}&=&\Big(\frac{g\mu_Bh}{2}-\frac{J_{\bot}}{2}-\frac{J_{\|}}{2}-\frac{J_{frust}}{2}-
\frac{J_c}{4}\Big)|1\rangle^+_n|0\rangle_{n+1}\nonumber\\
&+&\Big(\frac{J_c}{4}+\frac{J_{\|}}{2}-\frac{J_{frust}}{2}\Big)|0\rangle_n|1\rangle^+_{n+1}.
\end{eqnarray}

\section{Magnetization at low fields}

In the first order with respect to $D$ the perturbed ground state
is the following,
\begin{equation}
|0\rangle=|0\rangle_{r-d}-\sum_{n,k,j}\frac{^j\langle
k,tripl|H^{DM}_n|0\rangle_{r-d}}{E_{tripl}(k,j)}|k,tripl\rangle^j.
\end{equation}
Using the local formula,
\begin{equation}
H^{DM}_n|0\rangle_n=
(-1)^nD\Big(\frac{1}{\sqrt{8}}\cos\theta(|1\rangle_n^++|1\rangle_n^-)+\frac{1}{2i}\sin\theta|1\rangle^0_n\Big),
\end{equation}
and taking into account that $\sum_n(-1)^n{\rm
e}^{-ikn}=N\delta_{k,\pi}$ we obtain,
\begin{equation}
|0\rangle=|0\rangle_{r-d}-D\sqrt{N}\Big(\frac{\cos\theta}{\sqrt{8}}\sum_{j=\pm1}\frac{1}{E_{tripl}(\pi,j)}
|\pi,tripl\rangle^j+\frac{\sin\theta}{2iE_{tripl}(\pi,0)}|\pi,tripl\rangle^0\Big).
\end{equation}

As it follows from (16) the first order correction diverges as
$\sqrt{N}$. Nevertheless as it will be shown below the uniform and
staggered magnetizations defined as \cite{3},
\begin{eqnarray}
{\bf m}_u(h)&=&\langle0|{\bf S}_{1,0}+{\bf S}_{2,0}|0\rangle,\\
{\bf m}_s(h)&=&\frac{1}{2}\langle0|{\bf S}_{1,0}-{\bf
S}_{2,0}-{\bf S}_{1,1}+{\bf S}_{2,1}|0\rangle,
\end{eqnarray}
remain finite. This phenomena which is rather common for spin
systems was also mentioned in \cite{6} where the correct
perturbation theory based on cluster expansions was developed. Our
naive calculations may be reproduced by this approach.

Using (16) and the following formulas,
\begin{eqnarray}
({\bf S}_{1,n}+{\bf S}_{2,n})|0\rangle_n=0,\quad ({\bf
S}^z_{1,n}+{\bf
S}^z_{2,n})|1\rangle_n^j&=&j|1\rangle_n^j,\nonumber\\
{[}{\bf S}^x_{1,n}+{\bf S}^x_{2,n}\pm i({\bf S}^y_{1,n}+{\bf
S}^y_{2,n}){]}|1\rangle_n^j&=&\sqrt{2}|1\rangle_n^{j\pm1},\nonumber\\
{[}{\bf S}^x_{1,n}-{\bf S}^x_{2,n}\pm i({\bf S}^y_{1,n}-{\bf
S}^y_{2,n}){]}|0\rangle_n&=&\mp\sqrt{2}|1\rangle_n^{\pm1},\nonumber\\
({\bf S}^z_{1,n}-{\bf S}^z_{2,n})|0\rangle_n&=&|1\rangle^0_n.
\end{eqnarray}
one obtain,
\begin{eqnarray}
{\bf m}_u^x(h)&=&0,\nonumber\\
{\bf
m}_u^y(h)&=&-\frac{D^2\sin{2\theta}}{8E_{gap}}\sum_{j=\pm1}\frac{j}{E_{gap}-jg\mu_Bh},\nonumber\\
{\bf m}_u^z(h)
&=&\frac{D^2\cos^2\theta}{8}\sum_{j=\pm1}\frac{j}{(E_{gap}-jg\mu_Bh)^2},\nonumber\\
{\bf m}_s^x(h)&=&\frac{D\cos\theta}{2}\sum_{j=\pm1}\frac{j}{E_{gap}-jg\mu_Bh}\nonumber\\
{\bf m}_s^y(h)&=&0,\quad {\bf m}_s^z(h)=0.
\end{eqnarray}
For $h\ll h_c$ or equivalently $g\mu_Bh\ll E_{gap}$ these formulas
reduce to a compact form,
\begin{eqnarray}
{\bf m}_u({\bf h})&=&\frac{g\mu_B}{2E_{gap}^3}{[}{[}{\bf
D}\times{\bf
h}{]}\times{\bf D}{]},\nonumber\\
{\bf m}_s({\bf h})&=&\frac{g\mu_B}{E_{gap}^2}{[}{\bf D}\times{\bf
h}{]},
\end{eqnarray}
similar to the one obtained in \cite{3} for a single dimer.

Finitely we notice that though the perturbed ground state (16) is
expanded only up to the first order of $D$, the uniform
magnetization is proportional to $D^2$. In general this result
requires also a $D^2$ term in the expansion for $|0\rangle$
because a matrix element between this term and $|0\rangle_{r-d}$
would be of order $D^2$. However according to the first formula in
the Eq. (19) the latter term always vanishes. Therefore our result
is correct.

\section{Magnetization at high fields}

At high fields it is more convenient to study a deviation of the
uniform magnetization from its saturation value ${\bf
m}_u^{sat}=(0,0,1)$,
\begin{equation}
\Delta{\bf m}_u(h)={\bf m}_u^{sat}-{\bf m}_u(h).
\end{equation}
In components,
\begin{eqnarray}
\Delta{\bf m}_u^{x,y}(h)&=&-\langle0|{\bf S}^{x,y}_{1,0}+{\bf
S}^{x,y}_{2,0}|0\rangle,\nonumber\\
\Delta{\bf m}_u^z(h)&=&\langle0|1-{\bf S}^z_{1,0}-{\bf
S}^z_{2,0}|0\rangle.
\end{eqnarray}

Both $\Delta{\bf m}_u(h)$ and ${\bf m}_s(h)$ may be calculated in
the same way as in the low-field case. The perturbed ground state,
\begin{equation}
|0\rangle=|0\rangle_{sat}-\frac{D\cos\theta}{\sqrt{8N}}\sum_n(-1)^n\sum_{k}\frac{{\rm
e}^{-ikn}}{E_{singl}(k)}|k,singl\rangle^j,
\end{equation}
reduces to,
\begin{equation}
|0\rangle=|0\rangle_s-\frac{\sqrt{N}D\cos\theta}{\sqrt{8}E_{singl}(\pi)}|\pi,singl\rangle.
\end{equation}

Substituting this expression into (23) and using (19), we obtain,
\begin{eqnarray}
\Delta{\bf m}_u^x(h)&=&\Delta{\bf m}_u^y(h)=0,\quad
\Delta{\bf m}_u^z(h)=\frac{D^2\cos^2\theta}{8E_{singl}^2(\pi)},\nonumber\\
{\bf m}_s^x(h)&=&\frac{D\cos\theta}{2E_{singl}(\pi)},\quad {\bf
m}_s^y(h)={\bf m}_s^z(h)=0.
\end{eqnarray}

As a consequence of (26),
\begin{equation}
\Delta{\bf m}_u^z(h)=\frac{1}{2}({\bf m}_s^x(h))^2.
\end{equation}

As it follows from (26) polarization for $D\neq0$ reaches the
saturation value only asymptotically. Therefore the saturation
field $h_s$ strictly speaking has a sense only for the free
Hamiltonian $\hat H^0$ and is defined as the minimal field
satisfying all the inequalities in (8).

At a high magnetic field $h\gg h_s$ taking in account (12) one can
reduce the system (26) as follows,
\begin{eqnarray}
\Delta{\bf
m}^z_u(h)=\frac{D^2\cos^2\theta}{8g^2\mu_B^2h^2}\Big(1+2\frac{J_{\bot}+2J_{\|}+J_c}{g\mu_Bh}\Big),\nonumber\\
{\bf
m}^x_s(h)=\frac{D\cos\theta}{2g\mu_Bh}\Big(1+\frac{J_{\bot}+2J_{\|}+J_c}{g\mu_Bh}\Big).
\end{eqnarray}

\section{Conclusion}
In the first order with respect to the staggered DM term we
obtained analytical expressions for the uniform and staggered
magnetizations at low (21) and high (28) magnetic fields. The
corresponding magnetization experiment was reported in \cite{7}.
Although the presented plot well reproduces a global behavior of
the magnetic curve between $h_c$ and $h_s$ the representation in
the marginal regions $h\ll h_c$ and $h\gg h_s$ is rather crude for
comparison with the formulas (21) and (28). Therefore we refrain
from any estimations for $D$ basing on the data presented in
\cite{7}.

\end{document}